\documentclass[prd,aps,preprint,nofootinbib]{revtex4}
\usepackage{epsfig}
\usepackage{amsmath}

\makeatletter
\def\slash#1{{\mathpalette\c@ncel{#1}}} % TeXbook, bottom of p360
\makeatother

\newcommand\beq{\begin{eqnarray}}
\newcommand\eeq{\end{eqnarray}}

\def\pslash{\rlap/{\mkern-1mu p}}

\def\GFt{\widetilde{G}_F}

\begin{document}

\preprint{BNL-NT-07/46} \preprint{RBRC-702}

\title{On the Relation Between Mechanisms for\\ Single-Transverse-Spin
Asymmetries\\[1cm]}

\author{Yuji Koike}
\affiliation{Department of Physics, Niigata University, Ikarashi,
Niigata 950-2181, Japan}
\author{Werner Vogelsang}
%\email{vogelsan@quark.phy.bnl.gov}
\affiliation{Physics Department, Brookhaven National Laboratory,
Upton, NY 11973}
\author{Feng Yuan}
%\email{fyuan@quark.phy.bnl.gov}
\affiliation{Nuclear Science Division, Lawrence Berkeley National
Laboratory, Berkeley, CA 94720}\affiliation{RIKEN BNL Research
Center, Building 510A, Brookhaven National Laboratory, Upton, NY
11973 \\}
%\date{\today}

\begin{abstract}
Recent studies have shown that two widely-used mechanisms
for single-transverse-spin asymmetries based on either twist-three
contributions or on transverse-momentum-dependent (Sivers) parton
distributions become identical in a kinematical regime of overlap.
This was demonstrated for the so-called soft-gluon-pole and hard-pole
contributions to the asymmetry associated with a particular quark-gluon
correlation function in the nucleon. In this paper, using semi-inclusive
deep inelastic scattering as an example, we extend the study to the
contributions by soft-fermion poles and by another independent twist-three
correlation function. We find that these additional terms organize
themselves in such a way as to maintain the mutual consistency of
the two mechanisms for single-spin asymmetries.
%\vspace{10cm}
\end{abstract}

\maketitle

%{\bf I. Introduction.}
\section{Introduction}

Much progress has been made in recent years in the theoretical
understanding of single-transverse-spin asymmetries (SSAs) in
high-energy hadronic scattering. Traditionally, two mechanisms
for generating SSAs in Quantum Chromodynamics (QCD) had been
identified and used in the literature. One is based on ``asymmetric''
transverse-momentum-dependent (TMD) parton distributions or
fragmentation functions, in particular the so-called Sivers \cite{Siv90}
and Collins \cite{Col93} functions, respectively. For phenomenological
analyses of SSAs based on the TMD functions, see [3]. Here, a TMD factorization
is employed. The other mechanism makes use of twist-3 quark-gluon
correlation functions, either in the
nucleon \cite{Efremov,qiu,Kanazawa:2000hz,Eguchi:2006qz,Eguchi:2006mc} or in
fragmentation \cite{Koike:2002ti}, and of collinear factorization.

The recent progress came initially with the study of the factorization
and universality properties of TMDs and the related spin-dependent
cross sections in
Refs.~\cite{BroHwaSch02,Col02,BelJiYua02,BoeMulPij03,JiMaYu04,ColMet04}.
A highlight of these investigations has been the finding
that the Sivers functions are
not universal in the usual sense, but enter with opposite sign
in semi-inclusive deep-inelastic scattering (SIDIS) and the
Drell-Yan process. These studies were subsequently extended to
more complicated QCD hard-scattering~\cite{gl,QVY,CQ}, which
revealed a very complex general pattern of non-universality of
the Sivers functions.

In addition to this, recent work \cite{Ji:2006ub,Ji:2006vf,Ji:2006br}
(hereafter referred to as JQVY06)
has focused on the relation between the two mechanisms for SSAs
in a physical process. As examples, the SIDIS and Drell-Yan processes
were considered in a situation where one has two physical scales: the
virtuality of the photon, $Q$, and the transverse momentum of
the virtual photon $q_\perp$ (in Drell-Yan), or of the final hadron
$P_{h\perp}$ (in SIDIS). Then, as was argued in JQVY06, in the regime
$\Lambda_{\rm QCD}\ll q_\perp\
({\rm or}\ P_{h\perp})\ll Q$ both the twist-3 and the TMD formalism
should apply, and if both are consistent they should describe the
same physics and hence become identical. This indeed was verified by
performing the following two calculations: first the SSA was computed
within the twist-3 formalism, focusing only on the twist-3
quark-quark-gluon correlation function $T_F(x_1,x_2)$ (the so-called
Qiu-Sterman matrix element, with $x_1$ and $x_2$ the light-cone momentum
fractions of the two quarks),
and on the contributions from soft-gluon poles (SGPs) and hard poles (HPs)
in the corresponding hard-scattering function. Secondly, the Sivers
function was calculated to first order in perturbation theory, also
in terms of the $T_F(x_1,x_2)$ matrix element and from the SGP and HP
contributions. The result was inserted into TMD factorization formula
to obtain the SSA, which led to the same answer as that found in the
twist-3 approach for $\Lambda_{\rm QCD}\ll q_\perp\
({\rm or}\ P_{h\perp})\ll Q$. In this sense, the two formalisms
are unified, providing a unique framework for the SSAs in Drell-Yan and
SIDIS.

The twist-3 formalism for SSAs was recently revisited in much detail in
Refs.~\cite{Eguchi:2006qz,Eguchi:2006mc} (hereafter referred to as
EKT06). Several important issues were clarified and a more complete
formula for the SSA in SIDIS was derived. Gauge-invariance and
factorization of the twist-3 single-spin-dependent cross section
were proved to lowest order. Also the contribution from gluon
fragmentation was computed. According to EKT06,
there are two types of twist-3 quark-gluon correlation functions in the
nucleon that contribute to the single-spin-dependent cross section,
$G_F(x_1,x_2)$ and $\GFt(x_1,x_2)$. These are defined by
\begin{eqnarray}
& &\int {d\lambda\over 2\pi}\int{d\mu\over 2\pi}
e^{i\lambda x_1}e^{i\mu(x_2-x_1)}
\langle p\ S_\perp |\bar{\psi}_j(0)
{gF^{\alpha\beta}(\mu n)n_\beta}
\psi_i(\lambda n)|p\ S_\perp \rangle\nonumber\\
& &\qquad=
{M_N\over 4} \left(\pslash\right)_{ij}
\epsilon^{\alpha pnS_\perp}
{G_F(x_1,x_2)}+
i{M_N\over 4} \left(\gamma_5\pslash\right)_{ij}
S_\perp^\alpha
{\GFt(x_1,x_2)}+ \cdots,
\label{twist3distr}
\end{eqnarray}
where $M_N$ is the nucleon mass, $\psi$ is the quark field,
$n^\mu$ is a light-like vector ($n^2=0$) with $p\cdot n=1$,
$\epsilon^{\alpha pnS_\perp}=\epsilon^{\alpha\lambda\mu\nu}p_\lambda
n_\mu S_{\perp \nu}$ with $\epsilon_{0123}=1$,
and the transverse spin vector satisfies
$S_\perp^2 = -1$, $S_\perp \cdot p=S_\perp \cdot n=0$.
We have suppressed the gauge-links that connect the fields
along the lightcone and make the matrix element in
Eq.~(\ref{twist3distr}) gauge-invariant. We note that
$G_F(x_1,x_2)$ is related to $T_F(x_1,x_2)$ of JQVY06 by $T_F(x_1,x_2)=
- \pi M_N G_F(x_1,x_2)$. Two additional functions
could be defined from a matrix element similar to that in
Eq.~(\ref{twist3distr}) in which, however, the gluon field
strength tensor is replaced by a covariant derivative. As
was shown in Ref.~\cite{Eguchi:2006qz}, these functions are
not independent but may be expressed in terms of $G_F$, $\GFt$
and the twist-2 helicity distribution function.

The main features of the twist-3 spin-dependent cross section
for SIDIS derived in EKT06 can be summarized as follows.
(i) Both $G_F(x_1,x_2)$ and $\GFt(x_1,x_2)$ contribute;
(ii) the SSA may be generated by three kinds of propagator poles in the
hard-scattering function: SGPs, soft-fermion poles
(SFPs) and HPs. The SGPs are characterized by $x_1=x_2$
in the functions $G_F$ and $\GFt$. For the SFPs, one has $x_1=0$,
$x_2\neq 0$ (or vice versa), and for the HP contributions the two
quark momentum fractions are different and non-vanishing.
All three types of poles are present in the
contributions associated with $G_F$. Because of its anti-symmetry property,
$\GFt(x_1,x_2)=-\GFt(x_2,x_1)$, which follows from parity- and
time-reversal-invariance of QCD, $\GFt(x,x)\equiv 0$, and hence
$\GFt$ contributes only through SFPs and HPs.

In the light of the results of EKT06 it is important to extend the
studies of JQVY06. As we mentioned above, only the ``dominant''
SGP and HP contributions associated with $G_F$ were considered
in JQVY06 (for which EKT06 found agreement). The SFPs, the parts
involving $\GFt$, and the gluon fragmentation contributions were
not taken into account in JQVY06. For a complete study of the
consistency of the two formalisms for SSAs, it is important to
investigate if these additional contributions also organize
themselves in such a way that they either become suppressed at
$q_\perp\ll Q$, and hence negligible from the point of
view of TMD factorization, or can be absorbed into the TMD Sivers
function. For example, one immediate reason for concern might be the
fact that there are two independent twist-3 distribution functions,
$G_F$ and $\GFt$, while only one TMD Sivers function
appears in the TMD factorization formalism. In this paper, we will
address all these issues.

The rest of the paper is organized as follows. In Sec. II, we will first
re-examine the soft-fermion-pole contributions associated with
$G_F$ in the twist-3 mechanism for SIDIS. We will show that
there are more diagrams than considered in EKT06 that contribute to the SFPs,
and we will discuss the effects of these diagrams. We then turn to
the contributions associated with $\GFt$. After that, we present
the single-spin-dependent cross section in the regime
$P_{h\perp}\ll Q$, taking into account all $G_F$ and $\GFt$
contributions in both the quark/antiquark and gluon fragmentation channels.
In Sec. III, we will present the ``perturbative'' one-loop Sivers function
in the region $\Lambda_{\rm QCD} \ll k_\perp \ll Q$, including all
relevant contributions from $G_F$ and $\GFt$. Using this Sivers function,
the single-spin-dependent cross section in the TMD factorization approach
is shown to agree with that in the twist-3 approach. This establishes the
true consistency between the TMD and twist-3 formalisms in SIDIS and
verifies the original conclusion of \cite{Ji:2006br}. In Sec.~IV, we summarize
our results.

\section{Twist-3 calculation}

In EKT06, the SIDIS cross section was calculated for the
full kinematic region of transverse momentum, $P_{h\perp}\gg\Lambda_{\rm
QCD}$. The twist-3 single spin-dependent cross section was shown to have
the following azimuthal dependence:
\beq
d\Delta\sigma \sim \sin(\phi_h -\phi_S)\left[
\hat{\sigma}_1 + \hat{\sigma}_2\cos\phi_h +\hat{\sigma}_3\cos 2\phi_h\right],
\label{azimuth}
\eeq
where $\phi_h$ and $\phi_S$ are the azimuthal angles of
the hadron plane and the transverse spin vector of the nucleon,
respectively, measured from the lepton plane in a frame where the
initial proton and virtual photon are collinear. The leading contribution
associated with the Sivers functions gives rise only to the term
proportional to $\hat{\sigma}_1$ in (\ref{azimuth}).
We also note that the terms
associated with $\hat{\sigma}_2$ and $\hat{\sigma}_3$
in the twist-three cross section are
power-suppressed relative to the $\hat{\sigma}_1$-term in the small $P_{h\perp}/Q$
region~\cite{Eguchi:2006qz,Eguchi:2006mc}. This remains true also when
the SFP contributions are taken into account. In the following analysis,
we will focus only on the leading ($\hat{\sigma}_1$) contribution with
respect to $P_{h\perp}/Q$, which is the relevant contribution for
making contact with the TMD formalism.

\subsection{Soft-fermion-pole contributions associated with $G_F$}

As was shown in EKT06, both $G_F(x_1,x_2)$ and $\GFt(x_1,x_2)$ can
give rise to a SSA in SIDIS when combined with the phase from the
on-shell part of an internal propagator: \beq {1\over x_i \pm
i\varepsilon}=P{1\over x_i} \mp i\pi\delta(x_i), \eeq where $x_i$
($i=1$ or $2$) is the light-cone momentum fraction carried by one
of the quark lines as defined in (\ref{twist3distr}). The pole
part of this propagator leads to terms involving $G_F(x,0)$ and
$\GFt(x,0)$. The fact that one of the fermion lines carries no
longitudinal momentum is the reason why the associated
contributions are called soft-fermion-pole (SFP) contributions.
In this subsection, we will only discuss the contributions
associated with $G_F$. The case for $\GFt$ will be treated in the
next subsection; in any case the discussion is essentially the
same for $G_F$ and $\GFt$. The SFPs appear both in the quark and
the gluon fragmentation channels. In the following, we will
discuss these separately.

\subsubsection{Quark/antiquark fragmentation contributions}
\begin{figure}[t]
\begin{center}
\includegraphics[width=8.5cm]{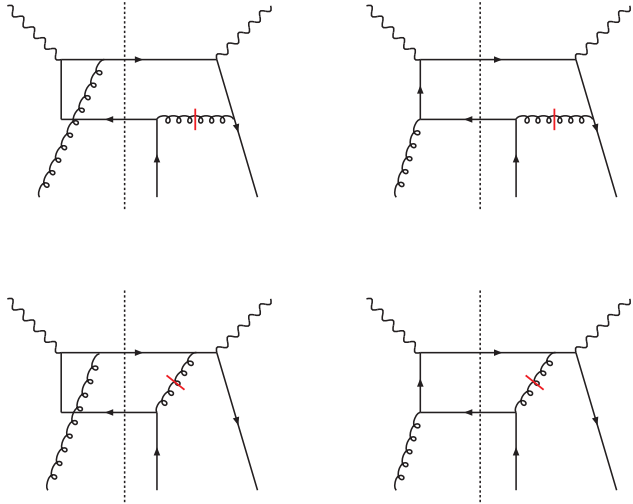}
\end{center}
\vskip -0.4cm \caption{\it Additional diagrams contributing to the
soft-fermion-pole contributions to the SSA in SIDIS for the
quark fragmentation channel. Their
contributions exactly cancel those of Fig.~6 of EKT06. Four more
diagrams are obtained by reversing the direction of the fermion
line and interchanging the momentum labels for the quark lines
(see below). For those diagrams, the lower of the two quark lines
going through the cut represents the fragmenting quark. \label{f1}}
\end{figure}
These contributions were calculated from the diagrams in Fig.~6
of EKT06, and the result for the $G_F$ contribution is given in
Eq.~(84) of their paper~\cite{Eguchi:2006mc}, and that for the
$\GFt$ contribution in Eq.~(92). Reversing the direction of the fermion
lines in these diagrams, one obtains the diagrams for the SFP
contributions in the antiquark fragmentation channel,
which were also included in the cross section formula of EKT06.

\begin{figure}[t]
\begin{center}
\includegraphics[width=8.5cm]{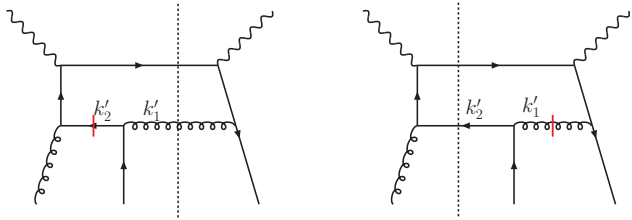}
\end{center}
\vskip -0.4cm \caption{\it An example of the cancellation between
the diagrams in EKT06 and those in Fig.~\ref{f1}. For the
quark fragmentation contribution, the two diagrams will
cancel because the soft-fermion attachment is to the ``unobserved''
final-state particle. The contributions of the diagrams for gluon or
antiquark fragmentation do not cancel, however, but are
power-suppressed in the limit $P_{h\perp}\ll Q$. } \label{f2}
\end{figure}

Careful scrutiny of the diagrams leads us to find two other sets of diagrams
that also give rise to SFP contributions. One set, for the
quark-fragmentation channel, is shown in Fig.~\ref{f1}. In these diagrams,
both the quark- and the antiquark line cross the final-state cut and
enter the nucleon matrix element on the right side of the cut. Despite
that, the diagrams give an SFP contribution to the cross section with
the same matrix elements as those in Fig.~6 of EKT06. They therefore
have to be included in the calculation.
It is not difficult to find that the contributions by the diagrams
in Fig.~\ref{f1} exactly cancel those from Fig.~6 of EKT06.
In order to better understand this cancellation, we redraw in Fig.~\ref{f2}
the upper right diagrams of Fig.~\ref{f1} and of Fig.~6 of EKT06. The
propagator giving the pole is indicated by a short bar in each case.
As one can see, these two diagrams are topologically identical, and the
only difference between them is the position of the final-state cut and
of the SFP. The scattering amplitude for the diagram on the left
has the following structure:
\begin{equation}
{\rm Left}:~~{\cal H}(x_1,x_2)\frac{1}{(k_2')^2+i\epsilon}\delta
((k_1')^2) \ ,
\end{equation}
where the momenta $k_1'$, $k_2'$ are as labelled in the figure.
For the right diagram, one has
\begin{equation}
{\rm Right}:~~{\cal H}(x_1,x_2)\frac{1}{(k_1')^2-i\epsilon}\delta
((k_2')^2) \ .
\end{equation}
When the propagator pole is taken, these two contributions
cancel. The other diagrams in Fig.~6 of EKT06 can be drawn in the
same manner, and one can check that they are each cancelled by the
corresponding diagram in Fig.~\ref{f1}.

We note that, in the two diagrams in Fig.~\ref{f2}, the soft-fermion
attaches to an ``unobserved'' final-state parton, that is, the
parton that does not fragment into the pion. It is instructive to recall
that the diagrams for the SGP contribution also cancel when
the soft-gluon-line is attached to an unobserved final parton
line\,\cite{qiu}. For the SGP this cancellation was clear, because
the two cancelling diagrams are just mirror images of each other,
with an opposite sign arising from the fact that the soft gluon
attaches on different sides of the cut. Here we have found that this
cancellation also occurs for the SFP contributions, despite the
fact that the two diagrams in Fig.~\ref{f2} are not mirror images
of each other.

A further set of diagrams giving rise to SFP contributions is
obtained by reversing the direction of the fermion lines in Fig.~\ref{f1},
along with an interchange of the assignment of the momenta for the external
fermion lines, so that now the other final-state line represents
the quark that fragments into the final pion. It turns out, however, that
this contribution is power-suppressed in the limit $P_{h\perp}\ll Q$.

The discussion for the antiquark-fragmentation channel is
completely parallel to that for the quark-fragmentation channel above.
As for quark fragmentation there are three types of diagrams, two of which
cancel, while the remaining one is power-suppressed for
$P_{h\perp}\ll Q$.

\subsubsection{Gluon fragmentation contribution}

For the gluon fragmentation channel, the diagrams in Fig.~6 of EKT06
(with appropriate momentum assignment) are the only ones that give
rise to SFP contributions. The corresponding partonic hard cross
sections are given in Eq.~(89) of~\cite{Eguchi:2006mc}
for the $G_F$ contribution, and in Eq.~(94) for the $\GFt$ one.
As it turns out, in the limit $P_{h\perp}\ll Q$, these contributions
are power-suppressed, as inspection of Eqs.~(89) and (94)
of~\cite{Eguchi:2006mc} shows. To obtain this result, one
uses the expansion
\begin{eqnarray}
\delta\left(\frac{\vec{q}_\perp^{\,2}}{Q^2}\right.&-&\left.
\left(1-\frac{1}{\xi}\right)
\left(1-\frac{1}{\hat \xi}\right)\right)\nonumber\\ &&\to
\left\{\frac{\hat \xi\delta(\xi-1)}{(1-\hat
\xi)_+}+\frac{\xi\delta(\hat
\xi-1)}{(1-\xi)_+}+\delta(\xi-1)\delta(\hat
\xi-1)\ln\frac{Q^2}{\vec{q}^{\,2}_\perp}\right\} \ , \label{e4}
\end{eqnarray}
for the phase-space delta function relevant for SIDIS,
which is valid for $\vec{q}_\perp^{\,2}\ll Q^2$. Here
$\vec{q}_\perp=-\vec{P}_{h\perp}/z_h$, $\xi=\hat x=x_B/x$
and $\hat \xi=\hat z=z_h/z$, as in EKT06. We note that
the first term in this expansion corresponds to contributions
that can be regarded as part of the TMD fragmentation function,
while the second term leads to terms to be absorbed into the
quark Sivers function. These observations will be useful later
when making contact with the TMD formalism. Substituting
the above expansion into the cross section formulas
given in EKT06, we straightforwardly find that the gluon fragmentation
terms from SFPs vanish in the limit $P_{h\perp}\ll Q$.

On the other hand, for gluon fragmentation there are SGP and HP
contributions associated with $G_F$ that survive in the limit
$P_{h\perp}\ll Q$. Their contributions can be derived from the
results in Eqs.~(87),(88) of~\cite{Eguchi:2006mc} by applying the
above expansion for the delta function.  We will include these
terms in our final results to be presented later.

\subsection{Contributions associated with $\GFt$}

A possibly important contribution not considered
in \cite{Ji:2006ub,Ji:2006vf,Ji:2006br} comes from the other
twist-3 distribution $\GFt(x_1,x_2)$ defined in (\ref{twist3distr}).
$\GFt$ is antisymmetric in its arguments~\cite{qiu,Eguchi:2006mc},
and hence $\GFt(x,x)\equiv 0$. Therefore, there are no SGP
contributions  to the single-spin-dependent cross section
associated with $\GFt$. However, as was shown in EKT06,
$\GFt$ does give rise to SFP and HP contributions. Regarding the
former, the situation in the quark/antiquark fragmentation
channel is the same as for $G_F$. After inclusion of the contributions
from the additional diagrams missing in EKT06, there is no surviving
SFP contribution in the $P_{h\perp}\ll Q$ limit.

The HP contributions for $\GFt$ come from the same diagrams
as for $G_F$; see~\cite{Ji:2006br} and Fig.~2
of~\cite{Eguchi:2006mc}. The results can be found in Eq.~(91)
of~\cite{Eguchi:2006mc} for the quark-fragmentation channel and in
Eq.~(93) for the gluon fragmentation channel. As for the $G_F$ case,
the contributions involve $\GFt(x_B,x)$, with one variable equal to
Bjorken-$x_B$. In the limit $P_{h\perp}\ll Q$ we use again
the expansion in~(\ref{e4}) and only consider the
leading-power contribution.  The first and the third terms of the
expansion do not contribute, because $\delta(\xi-1)$ means that $x=x_B$,
which leads to $\widetilde{G}_F(x_B,x_B)\equiv 0$.
Thus, we will only have contributions
from the $\delta (\hat \xi-1)$ term in~(\ref{e4}). These can be factorized
into the quark Sivers function.

\subsection{Twist-3 cross section in the $P_{h\perp}\ll Q$ limit}

We are now in the position to present the twist-3 single-spin-dependent
cross section in the regime $\Lambda_{\rm QCD}\ll P_{h\perp} \ll Q$,
including all the surviving contributions by $G_F$ and $\GFt$. It reads
\begin{eqnarray}
    \frac{d\Delta\sigma(S_\perp)}{dx_Bdydz_hd^2\vec{P}_{h\perp}}
      &=& -\frac{4\pi\alpha^2_{\rm
      em}S_{ep}}{Q^4}\epsilon_{\alpha\beta}S_\perp^\alpha
      \frac{z_hP_{h\perp}^\beta}{(\vec{P}_{h\perp}^2)^2}
      \frac{\alpha_s}{2\pi^2}
x_B\left( 1- y + {y^2\over 2}\right)
\int \frac{dxdz}{xz}\left\{D_q(z)\nonumber\right.\\
     && \left.\times \left[\delta(\hat \xi-1)A%\nonumber\right.\right.\\
%      &&\left.\left.
+\delta(\xi-1)B\right]+
C_F\delta(\xi-1)T_F(x,x)\,D_g(z)\frac{1+(1-\hat\xi)^2}{\hat \xi}\right\}\ ,
\nonumber\\
  \label{e7}
\end{eqnarray}
where
\begin{eqnarray}
A&=& \frac{1}{2N_C} \left\{ \left[x\frac{\partial}{\partial
x}T_F(x,x)\right](1+\xi^2)
      +T_F(x,x-\widehat{x}_g)\frac{1+\xi}{(1-\xi)_+}\nonumber\right.\\
 &&\left.     +T_F(x,x)\frac{(1-\xi)^2(2\xi+1)-2}{(1-\xi)_+}\right\}
+C_F T_F(x,x-\widehat{x}_g)\frac{1+\xi}{(1-\xi)_+}\nonumber\\
&&+\left(C_F+\frac{1}{2N_c}\right)\widetilde{T}_F(x-\widehat x_g,x)\ ,
\label{e8}\\
B&=& C_F T_F(x,x)\left[\frac{1+\hat\xi^2}{(1-\hat\xi)_+}
+2\delta(\hat\xi-1)\ln\frac{z_h^2Q^2}{\vec{P}_{h\perp}^2}\right] \
, \label{e9}
\end{eqnarray}
where $\widehat{x}_g\equiv(1-\xi)x=x-x_B$,
and we have used
the notation $\widetilde{T}_F(x_1,x_2)\equiv -\pi M_N \GFt(x_1,x_2)$
as well as ${T}_F(x_1,x_2)= -\pi M_N G_F(x_1,x_2)$
to make the comparison with Ref.~\cite{Ji:2006br} (Eq.~(18)) manifest.
$D_q$ and $D_g$ are the
usual transverse-momentum integrated
quark and gluon fragmentation functions, respectively.
Eq.~(\ref{e7}) extends the results in Eq.~(18) of \cite{Ji:2006br}
through the inclusion of the $\GFt$-term in~(\ref{e8}) and
the term involving $D_g$. The $B$ term in Eq.~(\ref{e9}) remains
the same as that in \cite{Ji:2006br}. This is crucial because
this term corresponds to a contribution by near-collinear gluon
emission from the unpolarized outgoing quark, which is completely
determined~\cite{JiMaYu04}. All other contributions in EKT06
are power-suppressed in the limit $P_{h\perp}\ll Q$.

To fully establish the consistency between the twist-3 and the
TMD factorization approaches in the intermediate region of the
transverse momentum, we now need to show that the result in
Eq.~(\ref{e7}) is reproduced in the TMD formalism. In particular,
one needs to address the issue that there are two independent
twist-3 distribution functions, $G_F$ and $\GFt$, in the
transversely polarized nucleon, while only one TMD Sivers function
appears in the TMD factorization. We will see in the next section
that there are also additional contributions to the Sivers function
that restore the equivalence of the two formalisms for
$\Lambda_{\rm QCD}\ll P_{h\perp} \ll Q$.

\section{Consistency with the TMD approach}

As was shown in~\cite{Ji:2006vf,Ji:2006br}, in the regime
$\Lambda_{\rm QCD}\ll k_\perp,p_\perp \ll Q$ the TMD quark Sivers
function $q_T(x,k_\perp)$ and the TMD quark fragmentation
function $\hat{q}(z,p_\perp)$ can be calculated perturbatively.
At the first order, one finds for the TMD quark fragmentation
function~\cite{Ji:2006br}:
\begin{eqnarray} \hat
q(z_h,p_\perp)&=&
\frac{\alpha_s}{2\pi^2}\frac{1}{\vec{p}_\perp^{\;2}}C_F\int\frac{dz}{z}
\left\{D_q(z) \left[\frac{1+\hat\xi^2}{(1-\hat\xi)_+}+\delta(\hat
\xi-1)
\left(\ln\frac{\hat\zeta^2}{\vec{p}_\perp^2}-1\right)\right]\right.\nonumber\\
&&+\left.D_g(z)\frac{1+(1-\hat\xi)^2}{\hat\xi}\right\}\ ,
\label{ffun}
\end{eqnarray}
where $D_q(z)$ and $D_g(z)$ are again the integrated quark and
gluon fragmentation functions. The $D_g$-term in this equation is
new compared to~\cite{Ji:2006br}; it comes from the radiation of a
gluon which then fragments.

Similarly, the TMD quark Sivers function $q_T(x,k_\perp)$ can be
calculated~\cite{Ji:2006vf}. The crucial point for our present
discussion is that it receives contributions not only from
$G_F$ (which were calculated in~\cite{Ji:2006vf}), but also by
$\widetilde G_F$. These can be calculated following the techniques
given in~\cite{Ji:2006vf}. The relevant diagrams are the same as
those in Fig.~10 of \cite{Ji:2006vf}, which were given there for the
HP contributions for $G_F$. One only needs to insert the appropriate
projection matrix for $\widetilde G_F$ as seen in Eq.~(\ref{twist3distr}).
Combining all contributions, one finds:
\begin{eqnarray}
q_T(x_B,k_\perp)=-\frac{\alpha_s}{4\pi^2}\frac{2M_N}
{(\vec{k}_\perp^2)^2}\int\frac{dx}{x}
\left\{A +C_FT_F(x,x)
\delta(\xi-1)\left(\ln\frac{x_B^2\zeta^2}{\vec{k}_\perp^2}-
1\right)\right\} \ , \label{sivpert}
\end{eqnarray}
where $A$ has been defined in Eq.~(\ref{e8}) and contains
the function $\widetilde G_F$.

Substituting the above expressions for the TMD quark Sivers function
and the TMD quark fragmentation function, along with the soft factor
calculated in~\cite{Ji:2006ub,Ji:2006vf,Ji:2006br}, into the TMD
factorization formula \cite{JiMaYu04}, one reproduces the
differential cross section in Eq.~(\ref{e7}) obtained from
the twist-3 quark-gluon correlation approach in the
intermediate-transverse-momentum region
$\Lambda_{\rm QCD}\ll P_{h\perp}\ll Q$. This demonstrates the full
consistency of these two approaches in SIDIS.

We close this section with two further observations. First, we emphasize
that the new contribution to the quark Sivers function associated with
$\GFt$ still obeys the well-known non-universality property of the quark
Sivers function. In particular, it will change sign for the Drell-Yan
process \cite{Ji:2006ub,Ji:2006vf,Ji:2006br}.

Secondly, the above result shows that the perturbative
calculation of $G_F$ receives contributions both from
$G_F$ and $\GFt$. This is a consequence of the fact that
$G_F$ and $\GFt$ mix under renormalization, as is well known
in the context of the renormalization of the twist-3 part of the
$g_2$ structure function which can be represented by the same two
functions\,\cite{Ratcliffe:1985mp}.

\section{Summary}

We have revisited the relation uncovered in~\cite{Ji:2006ub,Ji:2006vf}
between the twist-3 and the TMD formalisms for the single-transverse-spin
asymmetry in semi-inclusive lepton scattering. We have considered
additional contributions in the twist-3 formalism, which were recently
derived in~\cite{Eguchi:2006mc}. These contributions are
associated with soft-fermion poles, with another twist-3
correlation function, termed $\GFt$ in~\cite{Eguchi:2006mc},
and with gluon fragmentation. We have investigated the behavior
of these contributions in the limit $P_{h\perp} \ll Q$ and verified
that the new contributions organize themselves in a way consistent
with the transverse momentum dependent factorization formalism.
An important ingredient to this is that the quark Sivers function itself
receives contributions from the twist-3 quark gluon correlation function
$\widetilde G_F$, in addition to that from the previously considered
Qiu-Sterman matrix element $G_F$. This also indicates that $G_F$ and
$\GFt$ will mix under renormalization.

Thus, there is full equivalence of the two formalisms in
the intermediate-transverse-momentum regime. All our
calculations can be straightforwardly extended to the SSA in the
Drell-Yan process \cite{Ji:2006ub,Ji:2006vf}, and we find that
the consistency between the two approaches remains intact also here. \\

We thank Alessandro Bacchetta and Markus Diehl for bringing to our
attention the issue of consistency of the two formalisms in the
light of the soft-fermion-pole and $\GFt$ contributions derived in
EKT06. Y.K. thanks K. Tanaka for a useful discussion, and
F.Y. thanks X.N. Wang for a useful conversation. This work
was supported in part by the U.S.
Department of Energy under grant contract DE-AC02-05CH11231. F.Y.
and W.V. thank RIKEN, Brookhaven National Laboratory and the U.S.
Department of Energy (contract number DE-AC02-98CH10886) for
providing the facilities essential for the completion of their
work.

\end{document}